\documentclass[prd,showpacs,twocolumn,preprintnumbers,amsmath,amssymb,superscriptaddress,floatfix,nofootinbib]{revtex4-2}%

\usepackage{graphicx}
\usepackage{bm}
\usepackage{amsmath}
\usepackage{amsfonts}
\usepackage{amssymb}
\usepackage{color}
\usepackage{multirow}
\usepackage{subfigure}
\usepackage[colorlinks, citecolor=blue,anchorcolor=red,menucolor=red, linkcolor=red,filecolor=red,runcolor=red,urlcolor=blue,frenchlinks=red]{hyperref}

\begin{document}
	\title{Evidence of the open-flavor tetraquark  $T_{c\bar{s}2}$ in the process $B^+\to D^{*-}D_s^+\pi^+$}

	\author{Wen-Tao Lyu}
\affiliation{School of Physics, Zhengzhou University, Zhengzhou 450001, China}
	\vspace{0.5cm}

	\author{Li-Juan Liu}\email{liulijuan@zzu.edu.cn}
\affiliation{School of Physics, Zhengzhou University, Zhengzhou 450001, China}\vspace{0.5cm}

	\author{En Wang}\email{wangen@zzu.edu.cn}
\affiliation{School of Physics, Zhengzhou University, Zhengzhou 450001, China}\vspace{0.5cm}

\begin{abstract}
The newly observed open-flavor tetraquark $T_{c\bar{s}0}(2900)$ has attracted many attentions, and searching for its spin partners is crucial to exploring the internal structure of those states. In this work, we will show that, the $D_s^+\pi^+$ invariant mass distribution of the process $B^+\to D^{*-}D_s^+\pi^+$ measured by LHCb has a resonant-like structure around 2830~MeV, which could be associated with the predicted  $T_{c\bar{s}2}$, the spin $J=2$ partner of $T_{c\bar{s}0}(2900)$. Furthermore, we have evaluated the  momenta of the angular mass distribution, which are very different for each of the spin assumptions, and  have larger strength at the resonant energy than the peaks seen in the angular integrated mass distribution.
 We make a call for the experimental determination of these magnitudes, which could be used to pin down the existence of the $T_{c\bar{s}2}$.

\end{abstract}
	
	\date{\today}
	
	\maketitle
	
\section{Introduction}\label{sec1}

Since the charmonium-like state $X(3872)$ was observed by the Belle Collaboration in  2003~\cite{Belle:2003nnu}, many candidates of the exotic states were reported by experiments. In 2020, the LHCb Collaboration observed two charm-strange resonances $X_0(2900)$ and $X_1(2900)$ (renamed $T_{\bar{c}\bar{s}0}(2870)$
and $T_{\bar{c}\bar{s}1}(2900)$) in the process $B^+\to D^+D^-K^+$~\cite{LHCb:2020bls,LHCb:2020pxc}, which have several different theoretical interpretations about their nature, such as the compact tetraquark~\cite{He:2020jna,Karliner:2020vsi,Wang:2020prk,Yang:2021izl,Wang:2020xyc}, molecular structure interpretations~\cite{Wang:2021lwy,Chen:2020eyu,Lin:2022eau,Chen:2020aos,Huang:2020ptc,Liu:2020nil,Dai:2022qwh,Xiao:2020ltm,Hu:2020mxp,Molina:2020hde,He:2020btl}, or triangle singularity~\cite{Liu:2020orv}. Subsequently, the LHCb Collaboration analyzed the processes $B^0\to\bar{D}^0D_s^+\pi^-$ and $B^+\to D^-D_s^+\pi^+$ in 2022, and found two new open-flavor tetraquark states $T_{c\bar{s}0}(2900)^0$ and $T_{c\bar{s}0}(2900)^{++}$ in the $D_s^+\pi^-$ and $D_s^+\pi^+$ invariant mass distributions~\cite{LHCb:2022lzp,LHCb:2022sfr}. Since these two resonances contain four different flavor quarks $c\bar{s}\bar{u}d$~($c\bar{s}u\bar{d}$), and cannot be described within conventional quark models, 
there are several different interpretations for their structure, such as the `genuine' tetraquark state~\cite{Lian:2023cgs,Meng:2023jqk,Yang:2023evp}, the molecular state~\cite{Agaev:2022duz,Duan:2023lcj}.
Meanwhile, the $T_{c\bar{s}0}(2900)$ lies close to the thresholds of $D^*_s\rho$ and $D^*K^*$, thus the two-hadron continuum is expected to be of relevant for its existence, which makes the $T_{c\bar{s}0}(2900)$ natural candidate for the molecular state~\cite{Guo:2017jvc,Matuschek:2020gqe}.
In Ref.~\cite{Agaev:2022duz}, the authors argue that the $T_{c\bar{s}0}(2900)^{++}$ and $T_{c\bar{s}0}(2900)^0$ may be modelled as molecules $D_s^{*+}\rho^+$ and $D_s^{*+}\rho^-$, respectively. In addition, the $T_{c\bar{s}0}(2900)$ also can be considered as a virtual state created by the $D_s^*\rho$ and $D^*K^*$ interactions in coupled channels~\cite{Molina:2022jcd}, and the further analysis of the $D^*K^*$ interaction in a coupled-channel approach favors the $T_{c\bar{s}0}(2900)$ as a bound/virtual state~\cite{Duan:2023lcj,Lyu:2024wxa,Lyu:2023aqn,Lyu:2023ppb,Duan:2023qsg}.

Indeed, before the $X_0(2900)$ was reported by LHCb in Refs.~\cite{LHCb:2020bls,LHCb:2020pxc}, the $X_0(2900)$ and its spin partner states with $J=1,2$  were already predicted in Ref.~\cite{Molina:2010tx}.
Recently, Ref.~\cite{Molina:2022jcd} has studied $D^*K^*$ and $D_s^*\rho$ interaction within the local hidden gauge approach, and predicted a $J^P=2^+$ spin partner $T_{c\bar{s}2}$ with mass of 2834~MeV and width of 19~MeV. Meanwhile, a  $T_{c\bar{s}2}$ state with mass around 2800~MeV is also predicted in Ref.~\cite{Duan:2023lcj}. Thus, searching for the spin $J=2$ partner of $T_{c\bar{s}0}(2900)$ is crucial to testing the theoretical predictions of Refs.~\cite{Molina:2022jcd,Duan:2023lcj,Molina:2010tx}, and exploring the  internal nature of the open-flavor tetraquark $T_{c\bar{s}0}(2900)$.



Using proton-proton collision data at center-of-mass energies of $\sqrt{s}=7,8,13$~TeV, the LHCb Collaboration has measured the process $B^+\to D^{*-}D^+K^+$, where the $D^+K^+$ invariant mass distribution shows a peak structure around 2830~MeV~\cite{LHCb:2024vfz}, precisely the same mass of the predicted $T_{c\bar{s}2}$ of Ref.~\cite{Molina:2022jcd}. In Ref.~\cite{Lyu:2024zdo}, we have investigated this process by considering the $D^*K^*$ and $D_s^*\rho$ interactions, and found that the peak structure around 2830~MeV measured by LHCb could be associated with the predicted $T_{c\bar{s}2}$~\cite{Molina:2022jcd,Duan:2023lcj}. Furthermore, we have evaluated the momenta of the angular mass distribution of this  process, and shown that they are very different for each of the spin assumptions, which could be tested by experiments in future~\cite{Lyu:2024zdo}.
Subsequently, Ref.~\cite{Song:2024dan} has studied the $T_{c\bar{s}2}$ in the process $\Lambda_b\to\Sigma_c^{++}D^-K^-$ in a similar way, and predicted the signal of the $T_{\bar{c}s2}$ in the $D^-K^-$ invariant mass distribution.


Recently, the LHCb Collaboration has analyzed the process $B^+\to D^{*-}D_s^+\pi^+$, and reported that the fit fraction of the state $T_{c\bar{s}0}(2900)$ is less than 2.3\% at a 90\% confidence level~\cite{LHCb:2024vhs}. However, the $D_s^+\pi^+$ invariant mass distribution measured by LHCb has event excess around 2.83~GeV (see Fig.~9(b) of Ref.~\cite{LHCb:2024vhs}), which is consistent with the mass of the predicted $T_{c\bar{s}2}$~\cite{Molina:2022jcd,Duan:2023lcj}.


 
In the present work, we will continue with this line to investigate whether the open-flavor tetraquark $T_{c\bar{s}2}$, the spin partner of the $T_{c\bar{s}0}(2900)$, exists in the process $B^+\to D^{*-}D_s^+\pi^+$. Furthermore, we will predict the $D_s^+\pi^+$ moments of the invariant mass distribution from $l=0$ to $l=4$, which could be used to discriminate between the spins $J=0$, $1$, $2$ for the possible resonance. Given the compelling theoretical support for the $T_{c\bar{s}2}$ state, the work done here should be an incentive for this experimental analysis to be performed.

This paper is organized as follows. In Sec.~\ref{sec2}, we present the theoretical formalism of the process $B^+\to D^{*-}D_s^+\pi^+$. Numerical results and discussion are shown in Sec.~\ref{sec3}. Finally, we give a short summary in the last section.

\section{Formalism}\label{sec2}

\begin{figure}
	\centering
	\includegraphics[scale=0.65]{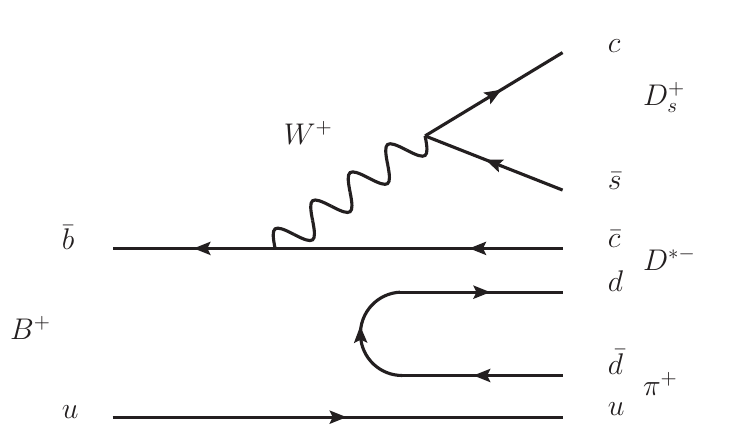}
	\caption{External emission mechanism of $B^+\to D^{*-}D_s^+\pi^+$.}
	\label{fig:quark-tree}
\end{figure}

\begin{figure}
	\subfigure[]{
  		\includegraphics[scale=0.65]{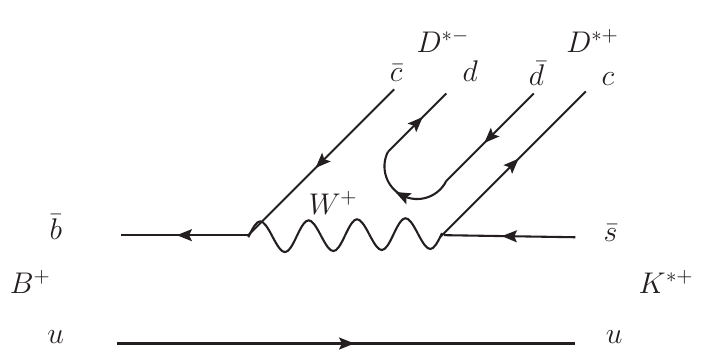}}
	\subfigure[]{
		\includegraphics[scale=0.65]{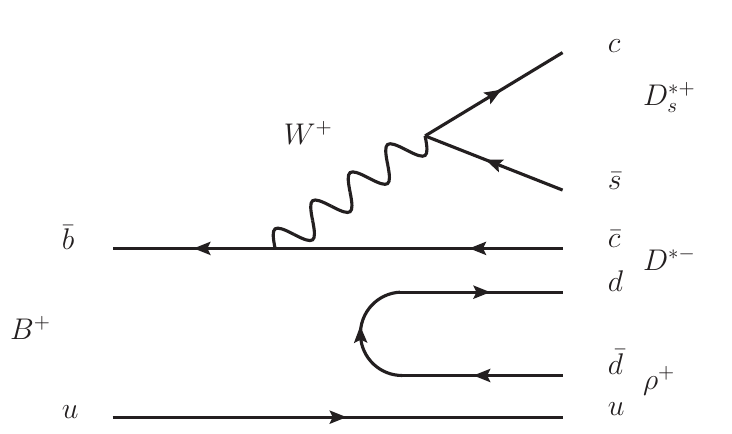}}
	\caption{Quark level diagrams for the processes $B^+\to D^{*-}D^{*+}K^{*+}$~(a) and $B^+\to D^{*-}D_s^{*+}\rho^+$~(b).}
	\label{fig:quark-interaction}
\end{figure}

In analogy to Refs.~\cite{Liu:2020ajv,Zhang:2020rqr,Dai:2018nmw,Wang:2017mrt,Li:2023nsw,Liu:2022dmm,Wei:2021usz,Zhang:2020rqr},
the process $B^+\to D^{*-}D_s^+\pi^+$ decay can proceed via the $W^+$ external emission, as shown in Fig.~\ref{fig:quark-tree}. The $\bar{b}$ quark from $B^+$ meson weakly decay into a $\bar{c}$ quark and a $W^+$ boson, and then the $W^+$ boson decays into a $c\bar{s}$ quark pair. Next, the $c\bar{s}$ quark pair will hadronize into $D_s^+$ meson. The $u$ quark of the $B^+$ meson and the $\bar{c}$ quark from $\bar{b}$ decay, together with the $d\bar{d}$ quark pair created from vacuum, hadronize into $D^{*-}$ and $\pi^+$ mesons. As a consequence, it is possible to directly produce $B^+\to D^{*-}D_s^+\pi^+$ at the tree level.

Since the predicted $T_{c\bar{s}2}$, as the spin partner of the $T_{c\bar{s}0}(2900)$, stems from the interaction of the $D^*K^*$ and $D_s^*\rho$ channels, we could also have the process $B^+\to D^{*-}D^{*+}K^{*+}$ via the $W^+$ internal emission mechanism and $B^+\to D^{*-}D_s^{*+}\rho^+$ decay via the $W^+$ external emission mechanism shown in Figs.~\ref{fig:quark-interaction}(a) and (b), following by transition of $D^{*+}K^{*+}/D_s^{*+}\rho^+\to D_s^+\pi^+$.

For the process $B^+\to D^{*-}D^{*+}K^{*+}$, considering that the angular momentum is conserved in the weak decay, $D^{*-}$ with $J^P=1^-$, $B^+$ with $J^P=0^-$, and $D^{*+}K^{*+}$ in $S$-wave, we need a $P$-wave mechanism to govern the reaction, contracting $\epsilon^\mu(D^{*-})$ with a vector, as discussed in Ref.~\cite{Lyu:2024zdo}. And for the process $B^+\to D^{*-}D_s^+\pi^+$, we want $D_s^+\pi^+$ coming from the $S$-wave $D^{*+}K^{*+}$ interaction, which will generate the $0^+$ or $2^+$ resonance, following by its decaying into $D_s^+\pi^+$ in $S$ or $D$-wave. Thus the $\epsilon^\mu$ polarization of the $D^{*-}$ has to be contracted with $P^\mu_{B^+}$. And the process $B^+\to D^{*-}D_s^{*+}\rho^+$ can proceed in a similar way.
In this case, 
following Refs.~\cite{Bayar:2022wbx,LHCb:2016lxy,Lyu:2024zdo,Song:2024dan,Wang:2021naf}, we can write the amplitude for the $B^+\to D^{*-}D_s^+\pi^+$,
\begin{equation}\label{Eq:1}
	\mathcal{T}=\epsilon_\mu(D^{*-})P^\mu_{B^+}(aY_{00}+bY_{20}+cY_{10}),
\end{equation}
assuming that we have first produced $D^{*-}D^{*+}K^{*+}~(D_s^{*+}\rho^+)$, and then the systems of the $S$-wave $D^{*+}K^{*+}$ and $D_s^{*+}\rho^+$ will transit into $D_s^+\pi^+$ in $S$-wave~($a$ term in Eq.~(\ref{Eq:1})), $D$-wave~($b$ term in Eq.~(\ref{Eq:1})). The $D^{*+}K^{*+}~(D_s^{*+}\rho^+)$ in $S$-wave cannot decay in $D_s^+\pi^+$ in $P$-wave, because of the parity conservation. However, it is very probable observed experimentally that a peak correspond to the $J^P=1^-$ state which couples to $D_s^+\pi^+$ in $P$-wave, and then we introduce the term of $bY_{20}$ to account for this possibility in Eq.~(\ref{Eq:1})~\cite{Lyu:2024zdo}. 

The invariant mass and angular distribution is given by
\begin{equation}
	\frac{d \Gamma}{d M_{\text{inv}}\left(D_s^+\pi^+\right) d \tilde{\Omega}}=\frac{1}{(2 \pi)^4} \frac{1}{8 M_{B^+}^2} p_{D^{*-}}\tilde{k} \sum|\mathcal{T}|^2,
\end{equation}
where $\tilde{\Omega}$ is the solid angle of $D_s^+\pi^+$ in their rest frame, and 
\begin{eqnarray}
	p_{D^{*-}}&=&\frac{\lambda^{1/2}\left(M_{B^+}^2, m_{D^{*-}}^2, M^2_{\text{inv}}(D_s^+\pi^+)\right)}{2M_{B^+}}, \nonumber\\
	\tilde{k}&=&\frac{\lambda^{1/2}\left(M^2_{\text{inv}}(D_s^+\pi^+), m_{D_s^+}^2, m_{\pi^+}^2\right)}{2M_{\text{inv}}(D_s^+\pi^+)},
\end{eqnarray}
with $\lambda(x,y,z)=x^2+y^2+z^2-2xy-2yz-2zx$.
We easily find for the sum over the $D^{*-}$ polarization in $|\mathcal{T}|^2$,
\begin{eqnarray}
	\sum|\mathcal{T}|^2 &=& \left( \frac{ M_{B^+}}{M_{D^{*-}}} \right)^2 p^2_{D^{*-}} \Big( |a|^2 Y_{00}^2 + |b|^2 Y_{20}^2 + |c|^2 Y_{10}^2 \nonumber\\
	&+& 2{\rm Re}(ab^*) Y_{00} Y_{20} + 2 {\rm Re}(ac^*) Y_{00} Y_{10}  \nonumber\\
	&+& 2{\rm Re}(bc^*) Y_{20}Y_{10}\Big),
\end{eqnarray}
then one can define the following moments,
\begin{eqnarray}
	\frac{d\Gamma_l}{dM_{\text{inv}}} = \int d \tilde{\Omega} \frac{d\Gamma}{dM_{\text{inv}} d \tilde{\Omega}} Y_{l0},
\end{eqnarray}
with the relation (see Eq.~(4.43) of Ref.~\cite{1957Elementary}),
\begin{eqnarray}
		 &&\int d \Omega Y_{l_3 m_3}^* Y_{l_2 m_2} Y_{l_1 m_1} \nonumber\\
		 &\quad=&\left[\frac{\left(2 l_1+1\right)\left(2 l_2+1\right)}{4 \pi\left(2 l_3+1\right)}\right]^{\frac{1}{2}} \mathcal{C}\left(l_1 l_2 l_3 ; m_1 m_2 m_3\right) \nonumber\\
		&& \times \mathcal{C}\left(l_1 l_2 l_3 ; 000\right),
\end{eqnarray}
where $\mathcal{C}\left(l_1 l_2 l_3 ; m_1 m_2 m_3\right)$ is the Clebsch-Gordan coefficients. Therefore, we can easily obtain the relations,
\begin{eqnarray}\label{Eq:gammal}
	\dfrac{d\Gamma_{0}}{dM_{\text{inv}} }&=&FAC \left[ \vert a \vert^{2} +\vert b \vert^{2} +\vert c \vert^{2} \right]  , \nonumber\\
	\dfrac{d\Gamma_{1}}{dM_{\text{inv}} }&=&FAC \left[ 2{\rm  Re}(ac^{*}) + \frac{4}{\sqrt{5}}{\rm  Re}(bc^{*}) \right]   , \nonumber\\
	\dfrac{d\Gamma_{2}}{dM_{\text{inv}} }&=&FAC \left[ \frac{2}{7}\sqrt{5} \vert b \vert^{2} + \frac{2}{5}\sqrt{5} \vert c \vert^{2} +2{\rm  Re}(ab^{*})\right]   , \nonumber\\
	\dfrac{d\Gamma_{3}}{dM_{\text{inv}} }&=&FAC  \sqrt{\frac{15}{7}} \times\frac{6}{5}{\rm  Re}(bc^{*})  , \nonumber\\
	\dfrac{d\Gamma_{4}}{dM_{\text{inv}} }&=&FAC ~\frac{6}{7} \vert b \vert^{2}  , 
\end{eqnarray}
where
\begin{equation}
	FAC =\frac{1}{\sqrt{4 \pi}} \frac{1}{(2 \pi)^4} \frac{1}{8 M^{2}_{D^{*-}}} ~ \vec{p}^{\,2}_{D^{*-}}~p_{D^{*-}}~ \tilde{k}. 
\end{equation}
Hence, the decay width could be related with the $\Gamma_0$ as
\begin{equation}
	\dfrac{d\Gamma}{dM_{\text{inv}} }=\sqrt{4\pi} \dfrac{d\Gamma_{0}}{dM_{\text{inv}} }.
\end{equation}
Here $d\Gamma_l/dM_{\rm inv}$ of Eq.~(\ref{Eq:gammal}) agree with  Refs.~\cite{Bayar:2022wbx,LHCb:2016lxy,Lyu:2024zdo,Song:2024dan,Wang:2021naf}.

In this work, we will take three cases for the total amplitude, in order to show the results for different spin assumption. First case is considering an $S$-wave resonance, $a \neq 0$, $b=c=0$, thus we only have the $d\Gamma_{0}/dM_{\text{inv}}$ term. Second case is considering a $P$-wave resonance, $a \neq 0$, $b=0$, $c \neq 0$, resulting in $d\Gamma_{0}/dM_{\text{inv}}$, $d\Gamma_{1}/dM_{\text{inv}}$, and $d\Gamma_{2}/dM_{\text{inv}}$ terms. Third case is considering a $D$-wave resonance (our preferred choice from present theoretical calculations),  $a \neq 0$, $b \neq 0$, $c = 0$ and we have $d\Gamma_{0}/dM_{\text{inv}}$, $d\Gamma_{2}/dM_{\text{inv}}$, and $d\Gamma_{4}/dM_{\text{inv}}$ terms.

\section{Results and Discussions}\label{sec3}

 With the above formalism, one can write down the amplitude from Eq.~(\ref{Eq:1}). For case I, we have,
\begin{equation}\label{Eq:case1}
	aY_{00}=\left(a_0 + a^\prime_0 \frac{M_{B^+}^2}{M^2_{\text{inv}}(D_s^+\pi^+)-M_R^2+i M_R \Gamma_R}\right) Y_{00},
\end{equation}
where $a_0$ is the contribution of tree level, and $a^\prime_0$ is the relative strength of the $S$-wave resonance. It should be stressed that, for the three cases, we assume $M_R=2834$~MeV and $\Gamma_R=19$~MeV, taken from Ref.~\cite{Molina:2022jcd}. We fit then $a_0$ and $a^\prime_0$ to the LHCb measurements~\cite{LHCb:2024vhs} from 2650~MeV to 3150~MeV, the energy region around 2834~MeV.

For case II, we have,
\begin{eqnarray}\label{Eq:case2}
	&&aY_{00}+cY_{10} \nonumber\\
	&&=a_1Y_{00} + c^\prime \frac{M_{B^+} \tilde{k}}{M^2_{\text{inv}}(D_s^+\pi^+)-M_R^2+i M_R \Gamma_R}Y_{10},
\end{eqnarray}
where we have put the factor $\tilde{k}$ suited to a $P$-wave amplitude. Here we consider the contribution  of the tree level in $S$-wave ($a_1Y_{00}$).  Then we fit again $a_1$ and $c^\prime$ to the LHCb measurements~\cite{LHCb:2024vhs}.

For case III, we have,
\begin{eqnarray}\label{Eq:case3}
	&&aY_{00}+bY_{20} \nonumber\\
	&&=a_2Y_{00} + b^\prime \frac{\tilde{k}^2}{M^2_{\text{inv}}(D_s^+\pi^+)-M_R^2+i M_R \Gamma_R}Y_{20},
\end{eqnarray}
where, again, we implement explicitly the factor $\tilde{k}^2$ suited to the $D$-wave resonance, and  consider the contribution  of the tree level in $S$-wave ($a_2Y_{00}$). We fit $a_2$ and $b^\prime$ to the LHCb measurements ~\cite{LHCb:2024vhs}.

 \begin{figure}[htb]
	\centering
	\includegraphics[scale=0.65]{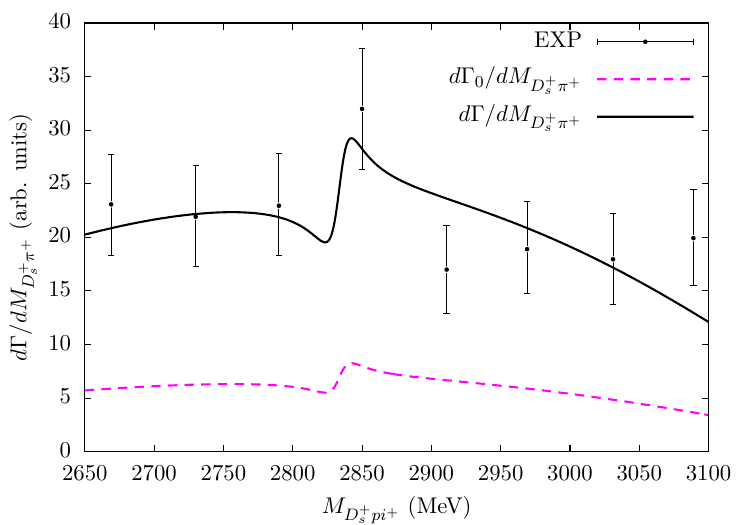}
	\caption{Fit to ${d\Gamma}/{dM_{D_s^+\pi^+}}$ in case I: $a \neq 0$, $b = c = 0$.}
	\label{Fig:CASE-I}
\end{figure}

\begin{figure}[htb]
	\centering
	\includegraphics[scale=0.65]{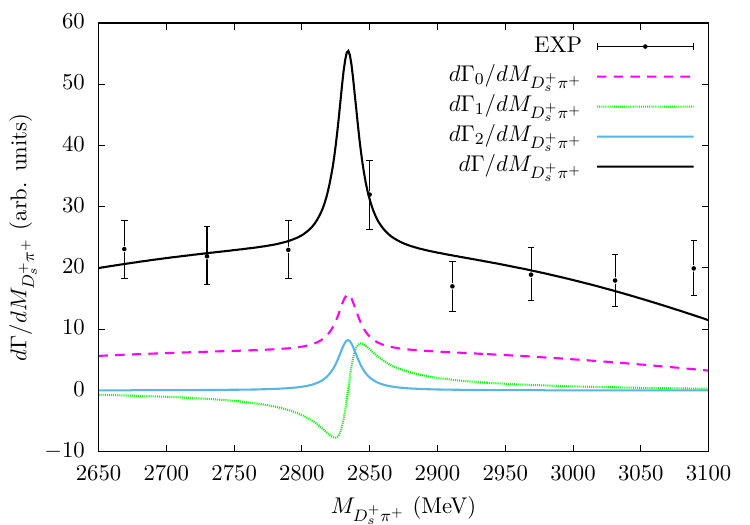}
	\caption{Fit to ${d\Gamma}/{dM_{D_s^+\pi^+}}$ for case II: $a \neq 0$, $b = 0$, $c \neq 0$.}
	\label{Fig:CASE-II}
\end{figure}

\begin{figure}[htb]
	\centering
	\includegraphics[scale=0.65]{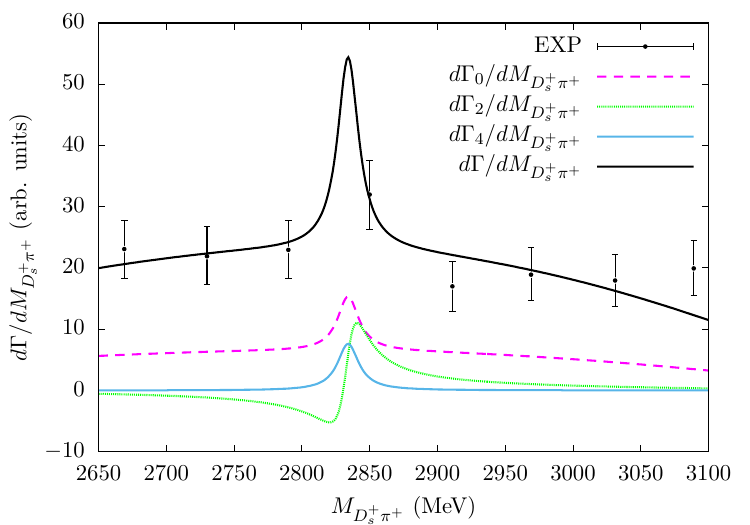}
	\caption{Fit to ${d\Gamma}/{dM_{D_s^+\pi^+}}$ for case III: $a \neq 0$, $b \neq 0$, $c = 0$.}
	\label{Fig:CASE-III}
\end{figure}

In the present work we follow the same path as in Ref.~\cite{Lyu:2024zdo}, and take a small background term $a_i$ ($i=1,2,3$) as follows,
\begin{equation}
	a_i \to \tilde{a}_i \frac{\tilde{k}}{M_B},
\end{equation}
where with this background we can fairly reproduce the LHCb measurement~\cite{LHCb:2024vhs}.

We show the results for case~I in Fig.~\ref{Fig:CASE-I}, and the fitted parameters are $\tilde{a}_0=8.93~\text{MeV}^{-1}$ and $a_0^\prime=4.84\times10^{-4}~\text{MeV}^{-1}$. The magenta-dashed curve shows the $D_s^+\pi^+$ invariant mass distribution of $d\Gamma_{0}/dM_{\text{inv}}$, and the black-solid curve shows the $D_s^+\pi^+$ invariant mass distribution  of $d\Gamma/dM_{\text{inv}}$. One should note that the $S$-wave background will interfere with the $S$-wave resonance, and because of $b = c = 0$, all of the other $d\Gamma_{l}/dM_{\text{inv}}=0$~($l\neq 0$). One can find that our results are in fair agreement with the LHCb measurements.

In Fig.~\ref{Fig:CASE-II}, we present the results for case~II, and we can also obtain the  acceptable fit parameters, $\tilde{a}_1=8.73~\text{MeV}^{-1}$ and $c^\prime=2.01\times10^{-2}~\text{MeV}^{-1}$. In this figure, the green-dotted curve shows the $D_s^+\pi^+$ invariant mass distribution  of $d\Gamma_{1}/dM_{\text{inv}}$, and the blue-solid curve shows the result of $d\Gamma_{2}/dM_{\text{inv}}$. One can see that the $d\Gamma_{1}/dM_{\text{inv}}$ is the interference between the contributions of the $S$-wave ($a_1$) and $P$-wave ($c'$), going from negative to positive, while the $d\Gamma_{2}/dM_{\text{inv}}$ is always positive. Our results of case II  are in agreement with the LHCb measurements.

In Fig.~\ref{Fig:CASE-III}, one show the results for case III with the $2^+$ resonance, and we can obtain the fitted parameters, $\tilde{a}_2=8.73~\text{MeV}^{-1}$ and $b^\prime=1.44\times10^{-1}~\text{MeV}^{-1}$. One can see that the green-solid curves show the $D_s^+\pi^+$ invariant mass distribution of $d\Gamma_{2}/dM_{\text{inv}}$, and the blue-solid curves show the result of $d\Gamma_{4}/dM_{\text{inv}}$. Our results of $d\Gamma/dM_{\text{inv}}$ are also in agreement with the LHCb measurements. Interestingly, our results indicate that, since the interference term  $d\Gamma_{2}/dM_{\rm inv}$ is linear in $b^\prime$, while $d\Gamma_{4}/dM_{\text{inv}}$ is quadratic in $b^\prime$, the strength of $d\Gamma_2/dM_{\text{inv}}$ is bigger than that of $d\Gamma_4/dM_{\text{inv}}$, or $d\Gamma_0/dM_{\text{inv}}$. In other words, the use of the momentum magnitude $d\Gamma_2/dM_{\text{inv}}$ has stressed the signal of the resonance versus the one obtained from $d\Gamma/dM_{\text{inv}}$.

We take advantage to mention, since the background term  does not interfere with the $P$-wave and $D$-wave resonance in the angle integrated mass distribution, then the $\tilde{a}_1$ and $\tilde{a}_2$ are the same. And we have taken $\tilde{a}_2$ versus $b^\prime$~(${a}_1$ versus $c^\prime$) of the same sign. If we reverse the relative sign of $\tilde{a}_2$ versus $b^\prime$~($\tilde{a}_1$ versus $c^\prime$), the interference magnitude can not change, but the line shapes of $d\Gamma_{1}/dM_{\text{inv}}$ for case~II and $d\Gamma_{2}/dM_{\text{inv}}$ for case~III will go from positive to negative.

One can see that, although the calculated $d\Gamma/dM_{\text{inv}}$ for each case are in agreement with the LHCb measurements, the predicted $d\Gamma_l/dM_{\text{inv}}$ are completely different for each of the spin assumptions, i.e., $J=0, 1, 2$. Thus, a careful study of the different $d\Gamma_{l}/dM_{\text{inv}}$ associated with the angular dependent mass distribution should be used to distinguish between  the different spins for the resonance.

\section{ Conclusions }
Recently, the LHCb Collaboration has analyzed the process $B^+\to D^{*-}D_s^+\pi^+$, and fitted the $T_{c\bar{s}0}(2900)$ state in $D_s^+\pi^+$ invariant mass distribution, but they do not find its clear signal~\cite{LHCb:2024vhs}. They, furthermore, cannot well describe the data in the $D_s^+\pi^+$ invariant mass distribution around 2830~MeV. On the other hand, one spin $J=2$ partner of the $T_{c\bar{s}0}(2900)$ is predicted to have a mass of 2834~MeV and a width of 19~MeV, which may be related to the structure around 2830~MeV in the $D_s^+\pi^+$ invariant mass distribution reported by LHCb.

Thus, we have carried out a detailed study of the possible signal of the open-flavor tetraquark state $T_{c\bar{s}2}$ in the $D_s^+\pi^+$ mass distribution around 2834~MeV. We assume three cases tied to three possible spins of resonance, which are $J=0$, $J=1$, and $J=2$, respectively.  We have calculated the angular integrated mass distributions and fit $d\Gamma/dM_{\text{inv}}$ to the data of LHCb~\cite{LHCb:2024vhs}, and found a good agreement with the LHCb data. One can find that the results of $d\Gamma/dM_{\text{inv}}$ alone cannot distinguish the spins $J=0,1,2$ of resonance. 

Furthermore, we have calculated the different momenta from $l=0$ to $l=4$, and show that they are drastically different for each of the spin assumptions, which could be used to distinguish the spin of resonance. We have also shown that, for $J=2$ case~($T_{c\bar{s}2}$), the momenta that involve interference of amplitudes have a strength bigger than the signal in the angular integrated mass distribution. This work has provided one method to confirm the existence of the predicted $J^P=2^+$ state $T_{c\bar{s}2}$, and  advocated the experimental determination of these magnitudes.

The fact that both the measurements on the processes $B^+\to D^{*-}D^+_
s \pi^+$ of Ref.~\cite{LHCb:2024vhs}  and $B^+\to D^{*-}D^+K^+$ of Ref.~\cite{LHCb:2024vfz} show a peak in the $D^+_s\pi^+$ or $D^+K^+$ invariant mass distributions at the same energy of 2830~MeV, precisely the same energy where the spin $J^P=2^+$ partner of the  $T_{c\bar{s}0}(2900)$ state predicted in Ref.~\cite{Molina:2022jcd} and close to the one predicted in Ref.~\cite{Duan:2023lcj}, gives a strong support to the idea that the peak observed in both reactions should correspond to that predicted state $T_{c\bar{s}2}$.  This should provide a sufficient motivation to undertake the task of determining the moments studied in the present work, where extra evidence for the existence of the resonance, and information on its spin could be obtained.

\section*{Acknowledgments}
We would like to acknowledge the fruitful discussions with Prof. Eulogio Oset. 
This work is partly supported by the National Key R\&D Program of China under Grant No. 2024YFE0105200, and by the Natural Science Foundation of Henan under Grant No. 232300421140 and No. 222300420554, the National Natural Science Foundation of China under Grant No. 12475086 and No. 12192263.

\end{document}